# Correctness Attraction: A Study of Stability of Software Behavior Under Runtime Perturbation

**Benjamin Danglot · Philippe Preux · Benoit Baudry · Martin Monperrus**

**Abstract** Can the execution of a software be perturbed without breaking the correctness of the output? In this paper, we devise a novel protocol to answer this rarely investigated question. In an experimental study, we observe that many perturbations do not break the correctness in ten subject programs. We call this phenomenon "correctness attraction". The uniqueness of this protocol is that it considers a systematic exploration of the perturbation space as well as perfect oracles to determine the correctness of the output. To this extent, our findings on the stability of software under execution perturbations have a level of validity that has never been reported before in the scarce related work. A qualitative manual analysis enables us to set up the first taxonomy ever of the reasons behind correctness attraction.

## 1 Introduction

In the introductory class of statics, a branch of mechanics, one learns that there are two kinds of equilibrium: stable and unstable. Consider Figure 1, where a ball lies respectively in a basin (left) and on the top of a hill (right). The first ball is in a stable equilibrium, one can push it left or right, it will come back to the same equilibrium point. On the contrary, the second one is in an unstable equilibrium, a perturbation as small as an air blow directly results in the ball falling away.

In one of his famous lectures [3], Dijkstra has stated that in software, *"the smallest possible perturbations – i.e. changes of a single bit – can have the most drastic consequences."*. Viewed under the perspective of statics, it means that Dijkstra considers software as a system that is in an unstable equilibrium, or to put it more precisely, that the correctness of a program output is unstable with respect to perturbations. However, previous works (eg [6, 10]) suggest the opposite, *i.e.* suggest that programs can accommodate perturbations.

In this paper, our goal is to empirically assess this hypothesis. We devise Attract, an experimental protocol to study the stability of program correctness under

University of Lille & Inria, France

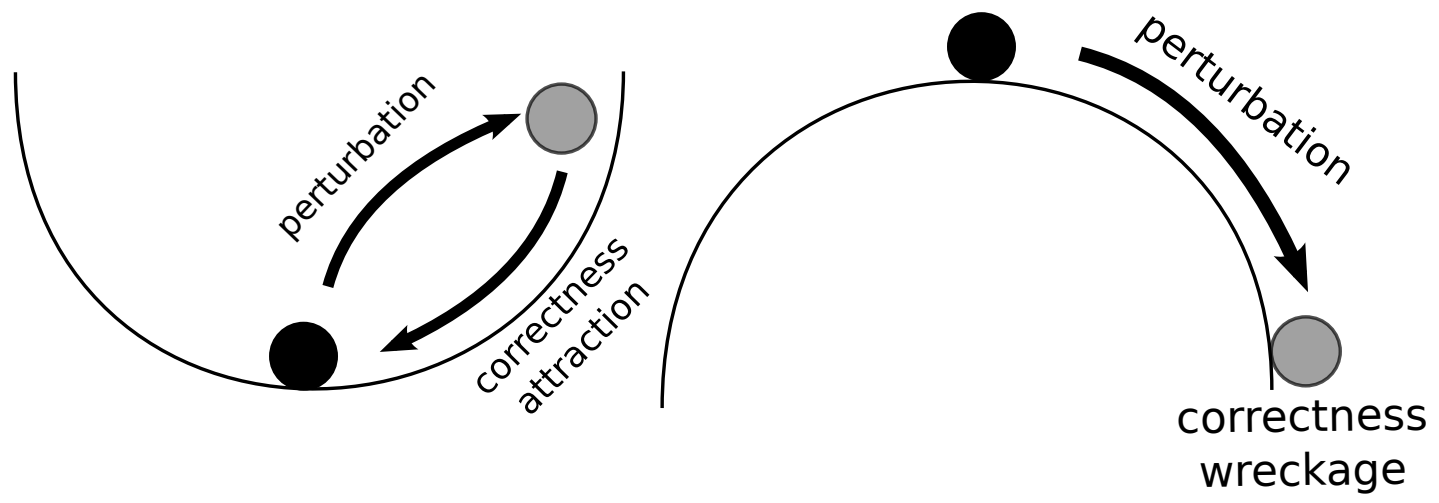

Fig. 1: The concept of stable and unstable equilibrium in physics motivate us to introduce the concept of "correctness attraction".

execution perturbation. Our protocol consists in perturbing the execution of programs according to a perturbation model and observing whether this has an impact on the correctness of the output. We shall observe two different outcomes: the perturbation breaks the computation and results in an incorrect output (unstable under perturbation), or the correctness of the output is stable despite the perturbation.

When a perturbation does not break output correctness, we observe "stable correctness", equivalent to stable equilibrium in statics. In such a case, we say that there is a phenomenon of "correctness attraction", an expression inspired by the concept of "attraction basin" in statics, which refers to the basin at the left hand-side of Figure 1, or more generally to the input points for which a dynamic system eventually reaches the same fixed and stable point.

We apply the protocol Attract to 10 Java programs ranging from 42 to 568 lines of code in two separate perturbation campaigns called PONE and PBOOL: the former perturbs integer expressions and the latter perturbs Boolean expressions. The comprehensive exploration of the perturbation space results in 2917701 separate executions of the ten programs. Among those 2917701 perturbed executions, 1977199 (67.76%) yield a correct output. This experiment suggests there is an important phenomenon of correctness attraction in software.

Next, we want to understand the reasons behind correctness attraction. We perform an in-depth analysis of a sample of program executions where correctness attraction occurs for all the ten considered programs (section 5). From this qualitative research, we set up a taxonomy (subsection 5.11) for correctness attraction, consisting of seven causes.

The key novelty of this research is systematization. The related work has only "loosely" observed correctness attraction because they have either considered weak correctness oracles or only perturbed a small number of executions. On the contrary, our protocol Attract is systematic in the sense that it only considers perfect oracles and performs a systematic exploration of a perturbation space.

Nonetheless, we note that those results are completely novel and that they are subject to bugs. Consequently, further experiments are required to validate them and to deepen our understanding of this intriguing phenomenon.

Our contributions are:

– A novel protocol for studying the perturbability of programs. This protocol yields a high level of validity because it considers perfect correctness oracles and a systematic exploration of the perturbation space for a given set of inputs.

- A large scale set of perturbability experiments over ten programs, and two perturbation models, resulting in 2917701 perturbed executions. We observe that 67.76% of perturbed executions yield a fully correct output. This means that correctness attraction is an important phenomenon in software.
- A qualitative manual analysis of 20 locations in the considered programs that are highly perturbable. From this analysis, we propose an original taxonomy of the reasons for correctness attraction, which discusses essential yet relatively under-researched properties of programs.

This document is organized as follows: in section 2, we define novel concepts for perturbation analysis. Then, in section 3 and section 4 we present two exhaustive empirical explorations of software perturbability. section 5 exposes a manual analysis of reasons behind correctness attraction. Finally, we present the related work in section 7 and we conclude in section 8.

## 2 The Attract Protocol

### 2.1 Definitions

We consider programs that take inputs and produce outputs. Each input can be characterized by an input model, for instance "all arrays of integers". The **correctness** of the output is checked by an **oracle**.

As observed by Barr and colleagues [1], there exists a large variety of oracle. Yet, we can distinguish between two main classes: partial oracles and perfect oracles. An oracle is said partial when it only checks for one aspect of the correctness: for example, if the program returns an array, a predicate that checks that only checks the length of the array is a partial oracle. An oracle is said perfect – a **perfect oracle** – when it fully assesses the correctness the output. For instance, a perfect oracle for a sorting algorithm asserts that the output is sorted, that all elements of the input array are still in the output array and that no other elements have been added. Using a reference implementation for the program under test is one approach to provide a perfect oracle.

In this paper, an **execution** refers to a pair (program, input), and we use the following core definitions.

**Definition 1** *An* ***execution perturbation*** *is a runtime change of the value of one variable in a statement or an expression. An execution perturbation has 3 characteristics:* time*: when the change occurs (e.g. at the second and the fourth iteration of a loop condition),* location*: where in the code (e.g. on variable 'i' at line 42) and* perturbation model*: what is this change, according to the type of the* location *(e.g. +1 on an integer value).*

**Definition 2** *The* ***perturbation space*** *for an input is composed of of all possible unique* ***perturbed executions*** *according to a perturbation model.*

We note that the size of a perturbation space depends on the perturbation model. For instance, when we perturb an integer value by adding to it in turn each integers between 0 and $2^{32}$ (a normal integer), the perturbation space is much larger than if we only increment the value. In this paper, our major goal is to exhaustively explore the perturbation space in order to have a really high internal validity. Consequently, we take a special care in devising perturbation models that imply spaces that are small enough to be systematically explored. In particular, we always perform a single execution perturbation per execution.

**Definition 3** ***Correctness attraction*** *is the phenomenon by which the correctness of an output is not impacted by execution perturbation. Correctness attraction means that one can perturb an execution while keeping the output correct according to a perfect oracle.*

### 2.2 Realization

To actually perform perturbations, we add **perturbation points** to the program under study where a **perturbation point** (denoted *pp*) is an expression of a given data type compatible with the perturbation model. For instance, if one perturbs integer values, the **perturbation points** will be all integer expressions (literal as well as compound expressions). In Listing 1[1], there are 3 potential integer perturbation points in a single statement, highlighted with braces.

$$\texttt{acc}\ \ |= \underbrace{\ i\ }_{pp_1} >> \underbrace{mask}_{pp_2};$$
$$\texttt{acc}\ \ |= \underbrace{i >> mask}_{pp_3};$$

Listing 1: Three integer **perturbation points** in a single statement.

In Attract, we statically locate **perturbation points** and automatically add perturbation code around them using a code transformation. The transformation consists of wrapping all compatible expressions into a function call $p$ (for "perturb")[2].

```
acc |= p(3, p(1,i) >> p(2,mask));
```

Listing 2: With perturbation code injected.

In Listing 2, each integer expression of Listing 1 is wrapped into a call to function $p$. Function $p$ takes two parameters: a unique index to identify the perturbation point and the expression being perturbed. If $p$ returns the second parameter, the transformed program is semantically equivalent to the original program. The identifier argument enables us to perturb only one location at a time. In our example, this identifier ranges from 1 to 3 corresponding to the index given in Listing 1 under perturbation point *pp*).

### 2.3 Example

Let us give the perturbation space for a more complex example shown in Listing 3, inspired from real code. Only one perturbation point is considered, when reading the value of $i$ in the loop statement.

---

[1] | is the bitwise or operator. >> is the binary right shift operator. The statement $acc \mid= i >> mask$ is equivalent to $acc = acc \mid (i >> mask)$.

[2] In our experiments, we implement this transformation on Java programs using the Spoon transformation library [9].

Table 2: Trace of three executions. The **+1** indicates when the perturbation on the value of $i$ occurs. The red value of *acc* indicates that the perturbation on $i$ has an impact on *acc* compared to the reference run. The green values signals that the value of *acc* again has the correct value.

| it. | *acc* | *i* | *acc* | *i* | *acc* | *i* |
|---|---|---|---|---|---|---|
| 1 | 2 | 8 **+1** | 2 | 8 | 2 | 8 |
| 2 | 3 | 7 | 2 | 7 **+1** | 3 | 7 |
| 3 | 3 | 6 | 3 | 6 | 3 | 6 **+1** |
| 4 | 3 | 5 | 3 | 5 | 3 | 5 |
| 5 | 3 | 4 | 3 | 4 | 3 | 4 |
| 6 | 3 | 3 | 3 | 3 | 3 | 3 |
| 7 | 3 | 2 | 3 | 2 | 3 | 2 |
| 8 | ③ | 1 | ③ | 1 | ③ | 1 |

```
int function(int bound) {
    int acc = 0;
    int mask = 0x2;
    for (int i = bound; i > 0; i--) {
        acc |= p(1, i) >> mask;
    }
    return acc;
}
```

Listing 3: Example of the instrumented function with the perturbation space reduced to a single perturbation point.

Consider the input *bound* = 8. After a first run of the program without any perturbation, one knows that perturbation point 1 is executed 8 times (the loop is bounded by the value of *bound*). Table 1 shows the trace of the perturbation-free execution – the reference run – for values *acc* and $i$. The circled value is the final output returned by the function: the value of variable *acc*.

Table 1

| it. | *acc* | *i* |
|---|---|---|
| 1 | 2 | 8 |
| 2 | 3 | 7 |
| 3 | 3 | 6 |
| 4 | 3 | 5 |
| 5 | 3 | 4 |
| 6 | 3 | 3 |
| 7 | 3 | 2 |
| 8 | ③ | 1 |

Then, we perform a systematic exploration of all possible perturbations. For the example, the perturbation model consists of increment integer expressions (called the PONE model in this paper). The traces of the perturbed executions are shown in Table 2, where a **+1** means that the perturbation has occurred at the given iteration. First, the exploration reveals the fact that the first three executions have the same final output as the reference run: 3. In the second execution the value *acc* is impacted by the perturbation, *i.e.* it is not equal to the value of the reference run at the same iteration (the red value). But at the next iteration, the computation makes it take again the right value (the green value), and the final output is indeed correct. For this example, there is a phenomenon of "**correctness attraction**".

Let us now consider more inputs, and not only *bound* = 8 and perform the exploration of the perturbation space for *bound* ranging from 0 to 100. We find out that over the 4950 executions required to explore the space, there are only 5 failures (the final output of the perturbed execution is different from the output of the reference run): $bound = 3$ at $it = 0$; $bound = 7$ at $it = 0$; $bound = 15$ at $it = 0$; $bound = 31$ at $it = 0$; $bound = 63$ at $it = 0$; This means that only 0.10% of the perturbation space yields an incorrect output. The 99.90% of perturbed yet correct executions indicates a high presence of **correctness attraction** in this code snippet for this particular perturbation point.

Table 3

| it. | *acc* | *i* |
|---|---|---|
| 1 | 1 | 3 **+1** |
| 2 | 1 | 2 |
| 3 | ① | 1 |

Let us discuss one of the 5 incorrect outputs due to perturbation, which happens when $bound = 3$ and when the perturbation occurs at the first iteration. Table 3 shows the corresponding execution trace. The perturbation occurs at the first iteration, where in the reference run $i = 3$ ($0011b$). When the perturbation occurs, then $i = 4$ ($0100b$) at the first iteration. Then, $0100b >> 2 = 0001b$ whereas in the reference run $0011b >> 2 = 0b$. The trace clearly shows that the computation does not produce a correct output.

## 2.4 The Core Perturbation Algorithm

We now formalize the example given in subsection 2.3 and present Algorithm 1 for exploring the perturbability of a program. It is based on the concept of perturbation points as presented in subsection 2.2. This algorithm requires: 1) a program 2) a perturbation model 3) a set of inputs 4) a perfect oracle.

The goal of this algorithm is to systematically explore the perturbation space. Algorithm 1 first records the number of executions of each perturbation point for each input in a matrix $R_{ref}$ (for reference run) without injecting any perturbation. $R_{ref}[pp, i]$ refers to the number of executions of perturbation point $pp$ for a given input $i$. Then, it re-executes the program for each input, with a perturbation for all points so that each point is perturbed at least and at most once per input. The oracle asserts the correctness of the perturbed execution (output $o$) for the given input ($i$). A perturbed execution can have three outcomes: a success, meaning that the correctness oracle validates the output; a failure meaning that the correctness oracle is violated (also called an oracle-broken execution); a runtime error (an exception in Java) meaning that the perturbation has produced a state for which the computation becomes impossible at some point (*e.g.* a division-by-zero). This algorithm performs a systematic exploration of the perturbation space for a given program and a set of inputs according to a perturbation model.

This algorithm enables us to compute:

- $\phi(pp)$, the proportion of correct executions per perturbation point ($s[pp]/\sum_i R_{ref}[pp, i]$);
- $\chi(pp)i$, the proportion of oracle-broken executions per perturbation point ($ob[pp]/\sum_i R_{ref}[pp, i]$);
- $\xi(pp)$, the proportion of exception-broken executions per perturbation point ($exc[pp]/\sum_i R_{ref}[pp, i]$);
- $\Omega$, the total number of correct executions over the whole perturbation space ($\sum_{pp} s[pp]$).
- $\Phi$, the percentage of correct executions called **correctness ratio** over the whole perturbation space ($\Omega/\sum_{pp}\sum_i R_{ref}[pp, i]$).

For a given program, if $\Phi$ is very low, Dijkstra's intuition is validated. If $\Phi$ is high, it means that the program under consideration exhibits a kind of correctness attraction.

**Input:**
*prog*: program,
*model*: perturbation model,
*I*: set of inputs for program *prog*,
*oracle*: a perfect oracle for program *prog*
**Output:**
*exc*: counters of execution per perturbation point,
*s*: counters of success per perturbation point,
*ob*: counters of oracle broken per perturbation point

```
instrument(prog)
for each input i in I do
    R_ref ← runWithoutPerturbation(prog, i)
    for perturbation point pp in prog do
        for j = 0, to R_ref[pp, i] do
            o ← runWithPerturbationAt(prog, model, i, pp, j)
            if exception is thrown then
                exc[pp] ← exc[pp] + 1
            else if oracle.assert(i, o) then
                s[pp] ← s[pp] + 1
            else
                ob[pp] ← ob[pp] + 1
            end if
        end for
    end for
end for
```

Algorithm 1: Systematic Exploration of the Perturbation Space. The statement *runWithoutPerturbation(prog, i)* returns a matrix ranging over perturbation points, and inputs: it contains the number of times each perturbation point is executed in the program *prog* for each input *i*. On the other hand, the statement *runWithPerturbationAt(prog, model, i, pp, j)* runs the program *prog* while using the perturbation model *model* the perturbation point *pp* at its $j^{\text{th}}$ execution for the given input *i*.

Note that $\Phi$ depends on the inputs: the larger the set of inputs being considered, the more diverse they are, and the more reliable is $\Phi$. Ideally, the considered inputs reflect the distribution of the production inputs.

## 2.5 Dataset

We study correctness attraction with a dataset of 10 programs. We have applied the following methodology to create this dataset: first and foremost, the programs can be specified with a perfect oracle; second, they are written in Java; third, they come from diverse application domains in order to maximize external validity. The resulting programs are summarized in Table 4. The first column displays the name of the program; the second is the number of lines of code; the third is a short description of the purpose of the program; the last column describes the perfect oracle used to evaluate the correctness of the output.

### *2.5.1 Representativeness of dataset*

We have taken a special care to have a representative dataset. First, the programs come from different application domains: mathematics, cryptography, biology, im-age processing. Second, the dataset contains a mix of 6 archetypal programs (*e.g.* quicksort) and 4 real programs used in production: RSA and RC4 are cipher algorithms from the widely-used library BouncyCastle. The implementation of Laguerre's method comes from one of the most widely used maths library in Java: Apache Commons Math. The linear regression program comes from the Weka library, developed by the machine learning group at the University of Waikato, and used in thousands of research projects.

Table 4: Dataset of 10 subjects programs used in our experiments. The phrase "reference output" refers to the expected output; the meaning should be clear from the context.

| Subject | LOC | Description | Oracle |
|---|---|---|---|
| quicksort | 42 | sort an array of integers | quicksort(x) = reference output |
| zip | 56 | compress a string with LZW | uncompress(compress(x)) = x |
| sudoku | 87 | solve a 9x9 sudoku grid | sudoku(x) = reference output |
| md5 | 91 | compute the MD5 hash of a string | md5(x) = reference output |
| rsa | 281 | RSA encrypt/decrypt with public and private keys | decrypt(encrypt(x)) = x |
| rc4 | 146 | RC4 encrypt/decrypt with symmetric key | decrypt(encrypt(x)) = x |
| canny | 568 | edge detector | canny(x) = reference output |
| lcs | 43 | compute the longest common sequence | lcs(x) = reference output |
| laguerre | 440 | find the roots of a polynomial functions | $\|poly(root)\| < 10^{-6} \| \forall root \in laguerre(poly)$ |
| linreg | 188 | compute the linear regression model for a set of points | $\|linreg(x) -$ reference coefficients$\| < 10^{-6}$ |

Beyond the representativeness of the programs, we also use representative inputs. For instance, for the lcs program (longest common subsequence), we have used real micro-RNA data coming from the mirBase biological database.

## 3 The PONE Experiment

We now present the PONE experiment. Its goal is to explore correctness attraction according to increments (+1) of integer values at runtime.

### 3.1 Perturbation model

In the PONE experiment, we perturb integer expressions. The PONE perturbation model is the smallest possible perturbation on an integer expression: a single increment of an integer value only once during an execution. An equivalently small perturbation model is MONE consisting of decrementing integers. As we will see later in subsection 3.4, MONE gives results that are very similar to those given by PONE.

### 3.2 Pilot experiment with QuickSort

We consider the implementation of *quicksort* exposed in subsection A.2. This implementation contains 41 integer expressions (integer-typed variables, integer literals, integer-typed compound expressions): these are all perturbation points for PONE.

We create 20 random inputs for *quicksort*: 20 random arrays of 100 integers between $-2^{31}$ and $2^{31}-1$. For these inputs, the 41 perturbation points are executed between 840 and 9495 times per input, which results in a perturbation space over the set of 20 inputs of 151 444 possible perturbations.

**Does *quicksort* exhibit correctness attraction under integer perturbation?** We have exhaustively explored the perturbation space according to the PONE perturbation model. Among the 151414 perturbed executions, 77% result in a perfectly correct output – a sorted array containing all input values in this case. This shows that *quicksort* is robust to execution perturbations. This result calls for further research questions: does this hold for other programs as well? Can we perturb other data types? Those research questions will be answered respectively in subsection 3.3 and section 4.

**In *quickSort*, are all perturbation points equally perturbable under PONE?** Table 5 provides the breakdown per perturbation point. The first column is the unique identifier of the perturbation point, the second is the number of executions in which a perturbation has occurred. The third, the fourth and the fifth columns are respectively the sum of the number of successes, oracle-broken executions and exceptions over all perturbed executions. Finally, the sixth and last column is the correctness ratio. We see that there is a great disparity in correctness ratio over perturbation points: 1) for three points, a single perturbation always breaks the output correctness (point #23, #35 and #40). We qualify this kind of points as **fragile** because a single perturbation at runtime breaks the whole computation. 2) some points can be systematically perturbed without any consequence on the correctness of the final output of the program (perturbation point #2 has a correctness ratio of 100%). We qualify this kind of points as **antifragile** (in opposition to fragile). The remainder is in between; those with a correctness ratio larger or equal than 75% are qualified as **robust**.

We will explore whether the first finding holds for other programs than *quicksort* in subsection 3.3 and we will perform a manual analysis of perturbation points to deepen our understanding of the second finding in section 5. Finally, we also see that the perturbation of a given perturbation point can result in both an invalid runtime state yielding an exception, and in an incorrect output as identified by the oracle (*e.g.* for point #39).

### 3.3 Generalization over 10 subjects

We now consider the PONE perturbation model over all ten subject programs of our dataset presented in subsection 2.5 and explained in appendix A.

Table 6 gives the results of the systematic exploration of the PONE perturbation space. For each subject, this table gives:

- the number of integer perturbation points $N_{pp}^{int}$;
- the number of perturbed executions (equal to the size of the PONE perturbation space);
- the number of **fragile** integer expressions;
- the number of **robust** integer expressions;
- the number of **antifragile** integer expressions;
- the **correctness ratio** (percentage of correct outputs) over all perturbed executions.

Table 5: The breakdown of correctness attraction per perturbation point in Quicksort for integer point.

| IndexLoc | #Perturb. Execs | $\phi$: #Success | $\chi$: #Failure | $\xi$: #Exception | $\Phi$: Correctness ratio |
|---|---|---|---|---|---|
| 0 | 1751 | 1202 | 543 | 6 | 68% |
| 1 | 1751 | 1641 | 0 | 110 | 93% |
| 2 | 1751 | 1751 | 0 | 0 | 100% |
| 3 | 1751 | 1751 | 0 | 0 | 100% |
| 4 | 1751 | 1751 | 0 | 0 | 100% |
| 5 | 1751 | 1751 | 0 | 0 | 100% |
| 6 | 1751 | 1751 | 0 | 0 | 100% |
| 7 | 1751 | 1751 | 0 | 0 | 100% |
| 8 | 1751 | 1751 | 0 | 0 | 100% |
| 9 | 1751 | 1739 | 0 | 12 | 99% |
| 10 | 5938 | 5938 | 0 | 0 | 100% |
| 11 | 5938 | 5938 | 0 | 0 | 100% |
| 12 | 8459 | 5946 | 2496 | 17 | 70% |
| 13 | 8459 | 8459 | 0 | 0 | 100% |
| 14 | 8459 | 8442 | 0 | 17 | 99% |
| 15 | 4272 | 4272 | 0 | 0 | 100% |
| 16 | 9495 | 6676 | 2691 | 128 | 70% |
| 17 | 9495 | 9477 | 0 | 18 | 99% |
| 18 | 9495 | 9495 | 0 | 0 | 100% |
| 19 | 5308 | 5308 | 0 | 0 | 100% |
| 20 | 4187 | 4187 | 0 | 0 | 100% |
| 21 | 4187 | 3616 | 571 | 0 | 86% |
| 22 | 3616 | 105 | 3506 | 5 | 2% |
| 23 | 3616 | 0 | 3564 | 52 | 0% |
| 24 | 3616 | 3616 | 0 | 0 | 100% |
| 25 | 3616 | 3616 | 0 | 0 | 100% |
| 26 | 1751 | 1633 | 118 | 0 | 93% |
| 27 | 1751 | 1751 | 0 | 0 | 100% |
| 28 | 840 | 275 | 565 | 0 | 32% |
| 29 | 840 | 840 | 0 | 0 | 100% |
| 30 | 1751 | 1751 | 0 | 0 | 100% |
| 31 | 1751 | 1632 | 119 | 0 | 93% |
| 32 | 891 | 321 | 570 | 0 | 36% |
| 33 | 891 | 801 | 0 | 90 | 89% |
| 34 | 3616 | 2361 | 1250 | 5 | 65% |
| 35 | 3616 | 0 | 3616 | 0 | 0% |
| 36 | 3616 | 1515 | 2101 | 0 | 41% |
| 37 | 3616 | 742 | 2822 | 52 | 20% |
| 38 | 3616 | 553 | 3058 | 5 | 15% |
| 39 | 3616 | 1420 | 2149 | 47 | 39% |
| 40 | 3616 | 0 | 3616 | 0 | 0% |

**Do the subjects exhibit correctness attraction?** As shown in Table 6, all subjects expose some level of correctness attraction. For instance, for *zip* with 20 inputs, the PONE systematic exploration comprises 38840 perturbed executions, of which more than 76% yield a correct output. The correctness ratio ranges from a minimum of 29% for *md5* to a maximum of 94% for *canny*. All programs are indeed perturbable according to PONE, and to a large extent. Recall, that we have built the benchmark with no prior analysis of its stability against perturbation: there is no positive bias in those results. In section 5, we provide a detailed account on the reasons for correctness attractions.

Table 6: PONE Results. The correctness ratio may not correspond directly to the number of Antifragile and Robust expressions because it is computed over all executions. Some points are executed much more than others, as explained in subsection 2.2.

| Subject | $N_{pp}^{int}$ | \|Search space\| | # Fragile exp. | #Robust exp. | # Antifragile exp. | $\Phi$: Correctness ratio |
|---|---|---|---|---|---|---|
| quicksort | 41 | 151444 | 6 | 10 | 19 | 77 % |
| zip | 19 | 38840 | 5 | 2 | 5 | 76 % |
| sudoku | 89 | 98211 | 12 | 27 | 8 | 68 % |
| md5 | 164 | 237680 | 102 | 24 | 7 | 29 % |
| rsa | 117 | 2576 | 55 | 8 | 32 | 54 % |
| rc4 | 115 | 165140 | 60 | 7 | 12 | 38 % |
| canny | 450 | 616161 | 58 | 129 | 133 | 94 % |
| lcs | 79 | 231786 | 10 | 47 | 13 | 89 % |
| laguerre | 72 | 423454 | 15 | 24 | 15 | 90 % |
| linreg | 75 | 543720 | 43 | 18 | 11 | 47 % |
| total | 1221 | 2509012 | 366 | 296 | 255 | 66 % |

**Are all perturbation points equally perturbable?** As shown in Table 6, all programs contain **antifragile** integer expressions. This ranges from 7 points for *md5* to 133 point for *canny*. Similarly, the number of robust integer expressions varies, raising as up as 129 points over 450 integer perturbation points for *canny* ($\approx$ 28% of all PONE perturbation points).

We now study the breakdown of perturbability per integer expression in the code. Figure 2 gives the distribution of the correctness ratio per perturbed expression as a violin boxplot. Each line represents a program, the violin represents the distribution. The white dot is the median of the distribution. For instance, the first row represents *quicksort*, and the fact that most of the distribution mass is at the right hand side reflects that most integer expressions are perturbable. For *quicksort*, this is a visual representation of the data shown in the last column of Table 5. We call this violin distribution the "perturbability profile" of a program.

We interpret Figure 2 as follows. First, there is no unique perturbability profile: for *quicksort*, most integer expressions are perturbable (first row, the mass is at the right hand side, that is around 100% of success); for *md5*, most integer expressions are not perturbable (fourth row, the mass is at the left hand side, that is around 0% of success); for *linreg*, there are two groups of points, fragile ones with low correctness ratio and antifragile ones with high correctness ratio. Second, this distribution is coherent with the large proportion of perturbed-yet-correct executions reported in Table 6. Since there is a large number of robust and antifragile integer expressions, they account for a large share of the PONE search space.

To sum up, the main conclusions of the PONE experiment are:

- The considered programs are perturbable according to the PONE perturbation model.

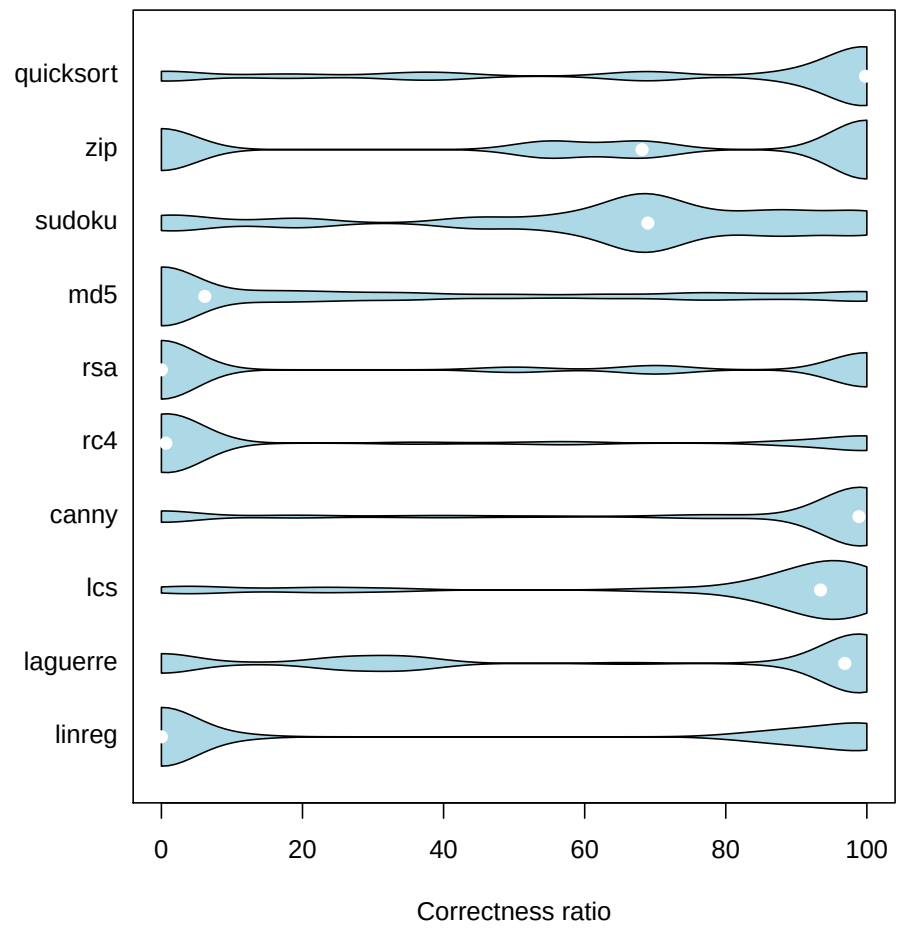

Fig. 2: Perturbability profiles for all subjects

- There are very few fully fragile integer expressions in the considered programs.
- There is a majority of highly perturbable integer expressions which results in a high level of correctness attraction.
- Dijkstra's view that software is fragile is not always true, correctness is rather a stable equilibrium than an unstable one.

### 3.4 The MONE Experiment

An other small integer perturbation is to decrement an integer expression: we call this perturbation the model MONE (Minus ONE). We have performed the systematic exploration of MONE. The results are really close to the results of PONE, and the presence of correctness attraction is similar. For instance, for *quicksort*, the correctness ratio under perturbation is 76,90% for MONE, very close to the 77.60% of PONE.

### 3.5 The PZERO Experiment

The PZERO perturbation model sets an integer value to zero at runtime. Similar to PONE and MONE, we have performed the systematic exploration of its corresponding perturbation space. On average, the correctness ratio for each program with PZERO is comparable, yet lower than the one obtained by PONE/MONE. For instance, we observe 63% of correctness attraction with PZERO and 77% with PONE.

Table 7: The breakdown of correctness attraction per boolean perturbation point in Quicksort.

| IndexLoc | #Perturb. Execs | $\phi$: #Success | $\chi$: #Failure | $\xi$: #Exception | $\Phi$: Correctness ratio |
|---|---|---|---|---|---|
| 0 | 5938 | 5932 | 0 | 6 | 99% |
| 1 | 8459 | 1779 | 6663 | 17 | 21% |
| 2 | 9495 | 1486 | 7991 | 18 | 15% |
| 3 | 4187 | 3616 | 571 | 0 | 86% |
| 4 | 1751 | 1087 | 664 | 0 | 62% |
| 5 | 1751 | 1072 | 679 | 0 | 61% |

## 4 The PBOOL Experiment

Now we consider another perturbation model: instead of perturbing integer expressions, we set up an experiment to perturb boolean expressions.

### 4.1 Perturbation model

The PBOOL perturbation model perturbs boolean expressions. One simple way to perturb a boolean expression is to flip the value from *true* to *false* and vice-versa, *i.e.* complement the original value.

An interesting feature of PBOOL is that it often results in changing the control flow, since boolean expressions acting as if or loop conditions are also perturbed. To this extent, a PBOOL perturbation can be considered as more radical than the PONE perturbation studied in section 3.

### 4.2 PBOOL on QuickSort

We consider the same implementation of Quicksort as presented in subsection A.2 and perform a systematic exploration of the perturbation space for the same 20 inputs. There are 6 boolean perturbation points in the considered implementation of Quicksort.

**Does Quicksort exhibit correctness attraction under boolean perturbation?** We have exhaustively explored the perturbation space of Quicksort for 20 inputs according to the PBOOL perturbation model. This results in 31581 perturbed executions, of which 47% result in a perfectly correct output. This is less than for PONE but still high. In the considered Quicksort implementation, all boolean expressions are in control-flow expressions (if or loop conditions). Consequently, all perturbations necessarily result in a different control flow. This 47% correctness ratio shows that in this program, many different execution paths exist that result in a fully correct output.

**In QuickSort, are all perturbation points equally perturbable under boolean perturbation?** Table 7 presents the results for all 6 boolean perturbation points in Quicksort. As we can see, there is also a variety in correctness ratios which ranges from 15% for boolean expression #2 to 99% for boolean expression #0. The 99% case will be studied in depth in subsection 5.1.

Table 8: The Results of the PBOOL Experiment.

| Subject | $N_{pp}^{bool}$ | \|Search space\| | # Fragile exp. | #Robust exp. | # Antifragile exp. | $\Phi$: Correctness ratio |
|---|---|---|---|---|---|---|
| quicksort | 6 | 31581 | 2 | 2 | - | 47.41 % |
| zip | 6 | 14280 | 5 | - | - | 0.78 % |
| sudoku | 26 | 28908 | 14 | 9 | 1 | 52.8 % |
| md5 | 10 | 12580 | 9 | - | - | 0.95 % |
| rsa | 20 | 620 | 7 | 2 | 5 | 49.68 % |
| rc4 | 7 | 10540 | 7 | - | - | 7.59 % |
| canny | 79 | 77038 | 28 | 10 | 14 | 71.55 % |
| lcs | 9 | 25165 | 6 | 1 | - | 55.13 % |
| laguerre | 25 | 109837 | 8 | 11 | - | 83.81 % |
| linreg | 15 | 98140 | 10 | - | 4 | 0.04 % |
| total | 203 | 408689 | 96 | 35 | 24 | 36.974 % |

### 4.3 Generalization over 10 subjects

We consider the same 10 subjects and oracles as for the PONE experiment. Table 8 gives the results of the PBOOL systematic exploration for the same 20 inputs. For each subject program this table gives:

- the number of boolean perturbation points;
- the number of perturbed executions (equal to the size of the PBOOL perturbation space);
- the number of fragile boolean expressions;
- the number of robust boolean expressions;
- the number of antifragile boolean expressions;
- the correctness ratio over all perturbed executions.

**Do the subjects exhibit correctness attraction under boolean perturbation?**

As shown in Table 8, 3 subjects (zip, md5, and linreg) are almost always broken under boolean perturbations, a single perturbation of a boolean expression in those programs breaks the output. For rc4, only 7% of outputs remain correct under boolean perturbations. On the other hand, for 7/10 subjects, the correctness ratio is high, from 37% to a maximum of 84% for laguerre. What we have observed for the small integer perturbation model PONE also holds for PBOOL: in certain programs, there exists multiple equivalent execution paths to achieve the same correct behavior. Although PBOOL perturbations are more radical than PONE perturbations, correctness attraction still occurs. In section 5, we perform a further manual qualitative analysis of this phenomenon.

**Are all boolean perturbation points equally perturbable?**

Figure 3 gives the distribution of the correctness ratio per perturbed boolean expression as a violin boxplot (the same visualization used and presented in section 3). Each line represents a program, the violin represents a mirrored distribution. The white dot is the median of the distribution. For Quicksort, this is another representation of the data shown in Table 7. Following the terminology set in subsection 3.3, those violin distributions are the “boolean perturbability profile” of programs. Here, we can see that there is no unique perturbability profile: for zip, no boolean expression is perturbable; for canny, many boolean expressions are perturbable.

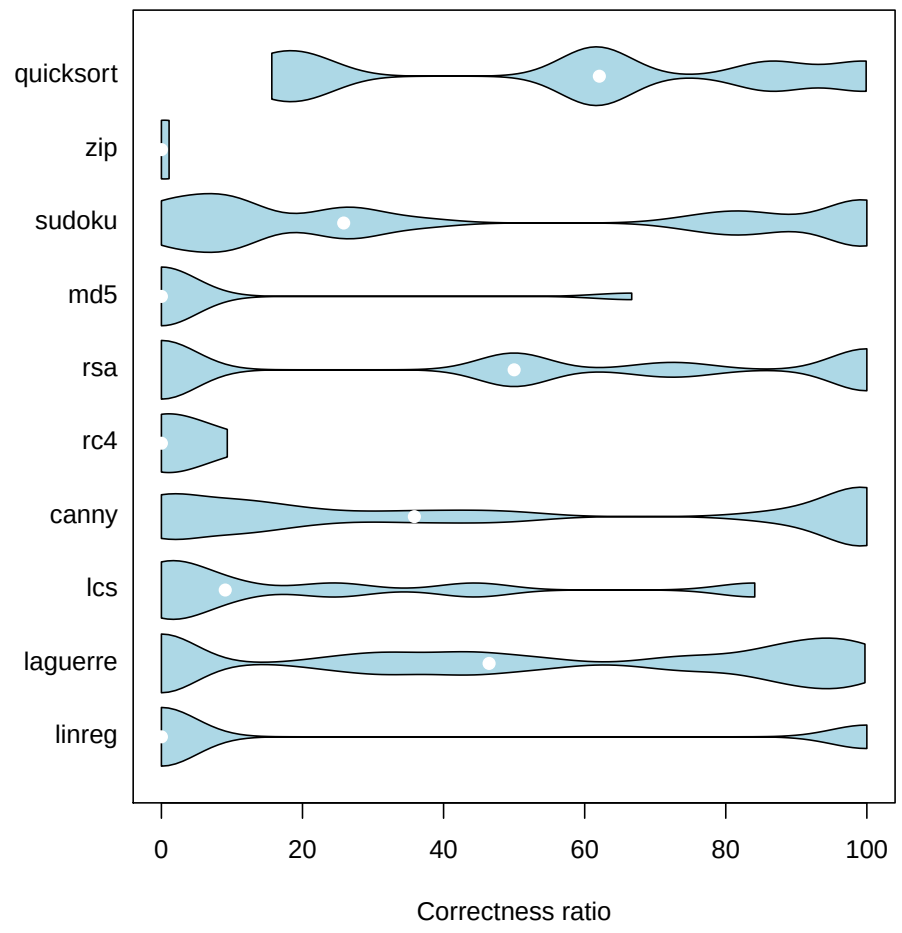

Fig. 3: The distribution of PBOOL Perturbability as Violin Distribution Plots.

To sum up, the main conclusions of the PBOOL experiment are:

– The observed perturbability of integer expressions also holds for boolean expressions: there is a phenomenon of correctness attraction that can happen for boolean perturbations.
– There are fully fragile subjects in the considered subjects.

## 5 Analysis of Correctness Attraction

**Research question: what are the reasons behind the correctness attraction phenomenon?**

For each subject and each experiment (PONE and PBOOL), we selected the perturbation points with the highest correctness ratio and manually analyzed them. For each subject, we present the most interesting integer perturbation point and the most interesting boolean perturbation point. This analysis enables us to setup a first taxonomy of the reasons behind correctness attraction.

The taxonomy will be presented in subsection 5.11. We think it is better to present it afterwards, because for such effects, it is more understandable to start with a concrete example. It also reflects the nature of this research: we started the analysis with no knowledge of the studied phenomenon.

### 5.1 Quicksort

Quicksort is a sorting program. The considered implementation sorts arrays of integers.

Integer Location 8:

$$\textbf{int}\ \texttt{pivot = array[}\ \underbrace{beg+((end-beg)/2)}_{\text{L8}}\ \texttt{];}$$

The L8 expression is used to select a value of the input array as pivot. It has a correctness ratio of 100%, which means that one can always select another pivot and still obtain a correctly sorted array. This fact is known: whatever the pivot used, the algorithm is still able to successfully perform the sorting task. Moreover, one can randomly select the pivot, and this randomization breaks the *worst case execution time* (*WCET*) of Quicksort, which is equivalent to the best practice of shuffling the array before running Quicksort [12].

Reason for correctness attraction: Natural randomization see 5.11.1.

Boolean Location 0:

$$\textbf{while}\ \texttt{(}\ \underbrace{(left <= right)}_{\text{L0}}\ \texttt{))}\ \{\ \ldots\ \}$$

The L0 expression controls a loop, we call it a *loop control point*. Under perturbation, either an iteration is added or the loop is stopped earlier. This perturbation point has 99.90% of correctness ratio. If the loop is stopped earlier than it should, there are more recursive calls, and the algorithm still converges to a correct state. The few failures happen when the perturbation adds an iteration.

Reason for correctness attraction: Fixed point effect see 5.11.3.

### 5.2 Zip

Zip is a compression program.

Integer location 0:

$$\textbf{int}\ \texttt{dictSize =}\ \underbrace{256}_{\text{L0}}\ \texttt{;}$$

This assignment it at the beginning of the decompress function of the LZW algorithm. By default, the algorithm has 256 entries to its dictionary. During compression, LZW adds 256 entries. With a PONE perturbation, $dictSize = 257$ while the actual size of the dictionary is 256 (from compression). Hence the $257^{th}$ entry remains empty. However, the rest of the algorithm never accesses this position, consequently, there are 100% of correct outputs after decompression.

However when we remove the last entry ($dictSize = 255$ under the MONE model), there are 6 tasks which fail. The reason is that they all contain the character ÿ which is encoded by the integer 255 (*i.e.* the last element of the dictionary), which is not added to the dictionary under MONE perturbation. When $dictSize = 255$, an access to the $256^{th}$ entry (at index 255) produces a failure. The phenomenon of a perturbation adding extra resources to the program is a common one, we call it the extra resource effect.

Reason for correctness attraction: Extra resources see 5.11.4

Boolean location 4:

$$\textbf{for}\ (\textbf{int}\ i = 0;\ \underbrace{i < 256}_{\text{L4}};\ i{+}{+})$$

```
    dictionary.put(i,""+(char)i);
```

The L4 expression controls a loop in method decompress too. It is the loop used to build the default dictionary, with the 256 characters already discussed. The perturbation stops the loop earlier or adds one iteration to it. When one iteration is added to the loop, it adds an extra yet useless entry to the dictionary. On the other hand, when the loop exits earlier, the dictionary lacks entries which results in a failure. This perturbation point has 1.09% of success, that is 56 successes over 5140 perturbed executions.

Reason for correctness attraction: Extra resources see 5.11.4

## 5.3 Sudoku

Sudoku is a solver of the Sudoku game.

Integer Location 78 and 79:

$$\text{mBoxSubset} = \textbf{new boolean}[\underbrace{mBoardSize}_{\text{L78}}][\underbrace{mBoardSize}_{\text{L79}}];$$

Expressions 78 and 79 are used in the initialization of the sudoku grid, *i.e.* mBoxSubset are the 9 boxes of 3x3 cells of the sudoku grids. The perturbation results with an extra row, column or box, which remain empty all along the computation of the solution because loops are bounded by the size of the input grid.

In contrast to PONE, a MONE perturbation on those locations results in 0 % of correctness ratio. As said, the loops are bounded by a fixed size, and the perturbation results in an out-of-bound access.

Reason for correctness attraction: Extra resources see 5.11.4

Boolean Location 97:

$$\text{setSubsetValue}(i,\ j,\ \text{value},\ \underbrace{true}_{\text{L97}});$$

Expression L97 is also in the initialization code. It is used to set up the constraints on the input grid, to say which numbers are present, in which rows, columns and boxes of the Sudoku grid. The last boolean parameter is used to specify whether the cell is empty or not. When it is set to “false” (which is what the perturbation does), the parameter “value” is simply ignored. It means that the perturbation results in a different input grid with one cell that is empty, and considered by the Sudoku solver as such. By construction, if the problem is satisfiable, it is also satisfiable with an easier one, and it admits more solutions. Consequently, the perturbation yields a high correctness ratio of 99.67%. However, in some rare cases, the solution found is different from the original one and this is considered as a failure.

Reason for correctness attraction: Relaxed problem see 5.11.5

### 5.4 MD5

MD5 is a hashing program.

Integer Location 3:

```
int numBlocks = (((messageLenBytes + 8) >>> 6) ) + 1);
                  \_______________________/
                            L3
```

This integer expression is at the beginning of method "compute()". The variable *numBlocks* is the number of blocks to be processed by the algorithm. When L3 with MONE, the perturbation is nullified by the subsequent bitshift because $messageLenBytes + 9 >>> 6$ is equal to $messageLenBytes + 8 >>> 6$. With a greater perturbation (for instance adding 200 to L3), the perturbation would not not nullified and the correctness of the output is broken.

Reason for correctness attraction: Nullified perturbation see 5.11.6

Boolean Location 22:

```
long messageLenBits = ((long) (messageLenBytes)) << 3;
for (int i = 0 ; i < 8 ; i++) {
                 \___/
                  L22
    paddingBytes[(((paddingBytes.length) - 8) + i)] = ((byte
        ) (messageLenBits));
    messageLenBits >>>= 8;
}
```

This is the only boolean perturbation point that gives a correctness ratio greater than 0%. It yields a high 66% correctness ratio. For this *loop control point*, the failure happens under perturbation when the loop does only one iteration or when it does one extraneous iteration. Interestingly, if the loop block is executed between 2-8 times, the result is correct. After analysis, the reason is that the other iterations are required, but only for longer inputs that the ones used in our experiments. Since we only use short inputs of 100 characters, the 2nd to 8th iterations can be safely skipped.

To verify our analysis, we a perturbed MD5 with a string of 10000 characters and the computation was indeed broken if the iterations were skipped due to the perturbation.

What we observe is this case is that the computed correctness ratio overfits our input distribution model as explained in section 6 (in this case, it overfits to strings of 100 characters).

Reason for correctness attraction: Overfit to input data see 5.11.7

### 5.5 RSA

RSA is a widely-used encryption algorithm (cf. subsection A.6).

Integer Location 0:

```
int bitSize = this.key.getModulus().bitLength();
              \____________________/
                        L0
if (forEncryption) {
```

```
    return (bitSize + 7) / 8;
} else {
    return (bitSize + 7) / 8 - 1;
}
```

The first statement assigns to *bitSize* the number of bits that are required to encode the actual value of the modulus of the secret key. The L0 expression, the BigInteger result of a method call, has a 100% correctness ratio with PONE. The value *bitSize* is used to check that the key is large enough to encrypt the current input. In our scenario, the modulus is encoded on 1024 bits. So, after the computation with PONE, we obtain 1025, and the bit length of 1025 is equal to the bit length of 1024. Consequently, there is no change in the returned value and the resulting correctness ration is 100%.

Reason for correctness attraction: Nullified perturbation see 5.11.6

Boolean Location 48:

```
if ( output.length < getOutputBlockSize() ) {
     \_____________________________________/
                    L48
    byte[] tmp = new byte[getOutputBlockSize()];
    System.arraycopy(output, 0, tmp, tmp.length - output.
        length, output.length);
    return tmp;
}
```

With PBOOL perturbations, we discover that perturbing the condition of this if-statement yields a 100% correctness ratio. Since this if-statement has no else branch, it means either not executing the else branch when it should or the opposite. In this case, the perturbation always flips false to true. This means that the code is not executed in nominal, non-perturbed mode, and executed in perturbed mode.

The then branch of this if statement copies the output array into a larger one in order to add a padding in front of the output array. When the perturbation occurs, the array is just copied, without any padding, which results in the exact same array. This results in extraneous memory usage, hence we classify it as Extra resources see 5.11.4.

Reason for correctness attraction: Extra resources see 5.11.4

## 5.6 RC4

RC4 is an encryption algorithm.

Integer Location 86:

```
for (int i=0; i < STATE_LENGTH; i++) {
              \_/
              L86
    engineState[i] = (byte)i;
}
```

Location L86 is 100% PONE-perturbable. When a PONE perturbation occurs, the loop body is executed one time less (the last iteration) and consequently, the last cell of the array is initialized to 0 (default value of byte) instead of $-1$.

This location is a *loop control point* as seen in Zip or MD5. This code is used to initialize the engine state in *setKey()* method. The perturbation results with a 0 (non initialized) instead of $-1$ in the last cell of the array. Later in the initialization (not shown here), this value is used in a bit swap, a xor and a mask (0xFF), for which 0 is equivalent to -1 for this data.

Reason for correctness attraction: Nullified perturbation see 5.11.6

Boolean Location 88:

$$\textbf{for}\ (\textbf{int}\ i=0;\ \underbrace{i < STATE_LENGTH}_{L88}\ )\ ;\ i{+}{+})$$

This is a *loop control point* with 6,23% of correctness ratio. This is also an initialization loop (same loop as in the integer location 86 that we have just discussed). Under perturbation, the loop is stopped earlier or does one more iteration than it should. An extra iteration always produces an error: an out of bound exception on the array engineState. However, the program still produces a correct output if the loop stop earlier, but not too early: as of the 237th iteration (until the 255th iteration) the loop can be stopped. We suspect that this is related to the size and diversity of the inputs that are created, but we lack domain-knowledge to strongly claim it.

Reason for correctness attraction: Overfit to input data see 5.11.7

## 5.7 Canny

Canny is an edge detector for pixel-based images.

Integer Location 3:

$$\textbf{return}\ \underbrace{Math.round(0.299f * r + 0.587f * g + 0.114f * b)}_{L3};$$

This perturbation point has 100% of correctness ratio. It computes the luminance of a pixel of the input image. The perturbation results with a stronger luminance for one given pixel. However, after that, there is a filter which averages the luminance by pack of 8 cells. Consequently, the perturbation evaprates, is nullified because it is not large enough to influence the average over 8 elements.

Reason for correctness attraction: Nullified perturbation see 5.11.6

Boolean Location 223:

$$\text{magnitude[index]}\ =\ \underbrace{gradMag >= MAGNITUDE_LIMIT?}_{L223}$$

```
MAGNITUDE_MAX : (int) (MAGNITUDE_SCALE * gradMag);
```

This perturbation point has 100% of correctness attraction. This conditional is used in order to trigger the “non-maximal suppression” of the canny filter. In our scenarios, the expression $gradMag >= MAGNITUDE_LIMIT$ is always false. So, the perturbation changes the value of the magnitude from $(int)(MAGNITUDE_SCALE* gradMag)$ to $MAGNITUDE_MAX$. This produces a failure when the real value of magnitude is $< 750$. However, for the 5 randomly selected inputs, this never happens. To verify this explanation, we perturbed this point with more inputs and we found that there are indeed inputs for which perturbing this location produces failure. To this extent, this is another case of overfitting.

Reason for correctness attraction: Overfit to input data see 5.11.7

### 5.8 LCS

Integer Location 2 and 5:

$$\textbf{int}\,[\,]\,[\,]\ \ \text{lengths} = \textbf{new}\ \textbf{int}\,[\underbrace{a.length()+1}_{\text{L2}}]\,[\underbrace{b.length()+1}_{\text{L5}}]\,;$$

Both locations have a 100% correctness ratio under PONE perturbation. The reason is that those perturbations result in extra allocated resources (as we have seen in Sudoku for example). The perturbations either add one row or column array `lengths`. Logically, a MONE perturbation results with a 0% correctness ratio.

Reason for correctness attraction: Extra resources see 5.11.4

Boolean Location 14:

```
for (int i = 0; i < a.length(); i++)
    for (int j = 0; j < b.length(); j++)
```

$$\underbrace{j < b.length()}_{\text{L14}}$$

This location has 44,36% of correctness ratio. It's a *control loop point*. Breaking the loop is equivalent to the well-known k-band optimization technique. It results in eliminating some values that are known worst than the one we are looking for. Consequently, when the optimization corresponds to an assumption that holds for the considered data, the output is correct. This happens in 44,36% of cases and we notice this happens if $i$ is near to 0 and $j$ near to $b.length()$(the left-bottom corner) or when $i$ is near $toa.length()$ and $j$ near to 0 (the right-top corner).

Reason for correctness attraction: Fixed point effect see 5.11.3

### 5.9 Laguerre

"Laguerre" is a numerical analysis program, it finds the roots of a polynomial equation.

Integer Location 73:

```
Complex[] coefficients = ...;
while (true) {
    ...
    pv = coefficients[ j ].add(z.multiply(pv));
    ...
}
```

$$\underbrace{j}_{\text{L73}}$$

This location has a correctness ratio of 99.28%. This loop is in method $solve()$, which is the core computation method.

The perturbation happens in a $while(true)$ loop. Each iteration of the loop slightly modifies the coefficients which are being computed. When a perturbation happens, one coefficient is not modified while the following is modified twice.

The endless loop is interrupted when a correct value is found. In other terms, the perturbation only delays the convergence to the desired value. This endless loop is a paradigmatic example of the fixed point effect on correctness attraction.

Interestingly, there are a few rare cases where the correctness is broken after the perturbation. This happens because the perturbation creates a fortuitous state where the loop-breaking condition evaluates to true.

Reason for correctness attraction: Fixed point effect see 5.11.3

Boolean Location 0:

```
if  (isSequence(min, z.getReal(), max))  {    // L0 = isSequence(min, z.getReal(), max)
    double tolerance = FastMath.max(getRelativeAccuracy() *
        z.abs(), getAbsoluteAccuracy());
    return (FastMath.abs(z.getImaginary()) <= tolerance) ||
        (z.abs() <= getFunctionValueAccuracy());
}
```

This location has 96.15 % correctness ratio under boolean perturbation. It is in the function *isRoot*() that checks if the argument is a root of the equation. First, if the perturbation occurs when the conditional is true, then the algorithm goes through the *else* branch, returns *false* and does a new iteration: the result is still correct because the next iteration finds the same real root (with tiny floating point difference). However, if the perturbation occurs when the actual conditional value is false, then the algorithm computes another condition. This other condition is a valid alternative check in most cases. It fails only for the last-but-one iteration, which explains the 3.85% of failures.

Reason for correctness attraction: Potential alternative executions see 5.11.2

## 5.10 Linreg

Linreg computes a linear regression using the Tikhonov regularization.

Integer Location 73:

```
Matrix x = new Matrix(a.getColumnDimension(), 1);    // L73 = 1
```

This location has a 100% correctness ratio. The perturbation results in an additional row to the Matrix x. This is an *extra resource* (as for sudoku). This point is in the method *aTy* which computes $a^t \times y$. The program changes the matrix to an array, and iterates over the length of the given matrix, consequently, the extra row is never used. To validate our hypothesis, we tried with MONE, and as we can expect, it results in 0% of correctness ratio, because the input x has a missing yet required row.

Reason for correctness attraction: Extra resources see 5.11.4

Boolean Location 27 :

```
do {
    ...
    try {
```

```
        solution = ssWithRidge.solve(bb);
        for (int i = 0; i < nc; i++)
            m_Coefficients[i] = solution.get(i, 0);
            success = true;
                      L27
    } catch (Exception ex) {
        ridge *= 10;
        success = false;
    }
} while (!success);
```

This location has 100% correctness ratio. It is a *control loop point* in a do-while. The perturbation results with an additional loop iteration. Instead of ending, it does a new call to the method solve(), since the computation is already at a stable fixed point, it does not break the correctness.

Reason for correctness attraction: Fixed point effect see 5.11.3

## 5.11 Taxonomy of the Reasons for Correctness Attraction

Thanks to our deep qualitative analysis of perturbable expressions, we are able to present a taxonomy of reasons for correctness attraction. This taxonomy is grounded on empirical observation: it is neither speculative, nor theoretical. Yet, it is probably not complete since our benchmark does not reflect the diversity of software. We expect future work on this topic to extend, and refine this taxonomy.

### *5.11.1 Natural randomization*

In a program, there may be locations where multiple values can be used to perform the same computation. For the sake of simplicity, optimization, or ignorance, the developer has hard-wired the use of one specific value. However, the alternative values are as valid as the hard-wired one.

Certain perturbations result in taking those alternative values, and correctness attraction is then explained by the presence of “natural randomization points” (where multiple valid values are legitimate) in the program.

### *5.11.2 Potential alternative executions*

In a program, there may exist different paths that are equivalent. In some cases, a perturbation triggers the execution of one such alternative path. This is different from the natural randomizability aforementioned which is about alternative values and not paths. In certain programs, the presence of alternative paths is one of the reasons explaining correctness attraction.

### *5.11.3 Fixed point effect*

Certain algorithms are based on the concept of fixed-point: they are designed to converge towards an expected correct value.

Certain perturbations result in slightly changing the convergence, for instance delaying it for some iterations. Consequently, we observe correctness attraction in programs which contain routines based on fixed-points.

#### *5.11.4 Extra resources*

Certain perturbations result in allocating extraneous resources (*e.g.* memory). They result in less efficient code but the functional correctness is not broken. However, we note that non-functional requirements on memory or execution time may be broken.

#### *5.11.5 Relaxed problem*

When a program solves a logical or a numerical problem, a perturbation may result in relaxing the problem under consideration. If the relaxation is small, it is likely that the solution to the relaxed problem is also a solution to the initial problem. This is one effect behind correctness attraction.

#### *5.11.6 Nullified perturbation*

By construction, all the considered perturbations have an effect on the program state. However, it happens that this effect is nullified (aka "masked" in the literature) by a subsequent operation. In this case, the computation quickly resumes a normal state after the perturbation, and correctness attraction is observed.

#### *5.11.7 Overfit to input data*

The last reason for correctness attraction is an artifact of our experimental protocol. We have observed cases where the correctness is kept only because the considered inputs did not trigger the broken behavior. This shows that the input generation model is important, and this is a threat to the validity of the outcomes of a perturbation experiment.

### 5.12 Potential Applications of Correctness Attraction

In this section, we discuss potential applications of correctness attraction. However, since we are the firsts to observe and reason about this phenomenon, those suggested applications are only tentative.

Execution randomization is one application of correctness attraction and antifragile point identification. Non-deterministic executions can increase the security of software systems, since they leak less knowledge to the attacker who aims at crafting an attack [2]. Our protocol to identify antifragile points is a way to engineer such randomization, because an antifragile point can be used to introduce a security-oriented randomization mechanism.

Second, another application of correctness attraction is black-box optimization. For instance, in the quicksort algorithm, our protocol enables one to identify the choice of pivot as a randomization point, which improves the performance on average and decreases the worst case complexity. Identifying such points with the protocol ATTRACT requires no domain knowledge, and hence can be called “blackbox optimization”.

Finally, we think that a deep understanding and mastering of correctness attraction can result in a significant breakthrough for software reliability. Let us consider the fact that a bug always starts with an unanticipated perturbation of the runtime state. Then, if we engineer techniques to automatically improve correctness attraction, we obtain zones that accommodate more perturbations of the runtime state, and those zones could be deemed “bug absorbing zones”.

## 6 Threats to Validity

We now discuss the threats to validity of our findings.

The first threat is a bug in the software written to perform this experiment. To mitigate this threat, we have used unit testing and continuous integration. Our code is publicly available at `https://github.com/Spirals-Team/jPerturb-experiments` for other researchers to verify and expand our results.

The second threat is about external validity. It may be that our focus on Java has an impact on the presence and prevalence of correctness attraction. Similarly, our benchmark of ten programs may also over- or under-estimate this effect. This would be fortuitous since we chose the subject programs at random.

Third, as discussed in subsubsection 5.11.7, we have discovered the presence of an overfitting phenomenon. This does not break our qualitative understanding of correctness attraction. However, it means that all the reported numerical figures may be an overestimation of the actual prevalence of correctness attraction.

## 7 Related Work

The related work on this topic is scarce.

Morell *et al.* perform “perturbation analysis” of computer programs in [8]. Their goal is completely different from ours: they want to evaluate the quality of a test strategy (akin to mutation testing). Consequently, they did not discover the phenomenon of correctness attraction. More generally, between mutation testing and runtime perturbations are of different nature: a classical mutation is a permanent change to the code, while a perturbation is a transient change to the state.

Wang *et al.* [15] studied the behaviour of programs when one forces them to take alternative branches. They instrumented 1000 conditional branches to force a program to take the alternate path (*i.e.* take the *else* path when it should take the *then* one and vice-versa). This forced path change is done only once per execution. In their experiments, they observe that in 50% of cases, this does not affect the program behaviour. They consider a perfect oracle: they compare the output of the perturbed execution against a reference output. Our results of the PBOOL experiment confirm their results, since many PBOOL perturbations effectively result in changing the executed branch. However, our results obtained by perturbing loop conditions and integer expressions go far beyond their experiments.

Rinard *et al.* [10] propose to shift our attention from perfectly correct behavior to acceptable behavior. To study acceptable behavior, they inject errors as follows: they change all the conditions of for loops by adding or removing one iteration: change $q > expr$ to $q \geq expr$ for instance. They explored the "acceptability envelopes" of two applications: pine a email client, and the Sure-Player MPEG decoder. The results show that while some injected errors are unacceptable, there are also some perturbations that are tolerated by the applications. Our work is along the same line yet differs significantly: first, our perturbation models are different; second, we exhaustively explore the perturbation space while they only explore a tiny fraction of it based on a manual usage of the perturbed software; finally, the correctness attraction we consider in this paper is a stronger notion of correctness than acceptable correctness.

In [6], X. Li and D. Yeung studied the impact of bit-flips on application-level correctness. They find that most perturbed executions produce an acceptable output for the user over 6 multimedia and AI benchmarks. This work is one of the most important one that considered the results of a traditional hardware fault injection model under the perspective of correctness. However, they go more towards acceptability-oriented computing than full correctness with perfect oracles as we do in this paper.

In [4], P. R. Eggert and D. S. Parker presents Wonglediff. Wonglediff is used to change the rounding modes of floating-point numbers used in a program. They use Wonglediff to analyze the portability of an application when the floating-point rounding mode changes (this indeed varies between OS's and machines). While the meaning of "perturbation" is close to what we consider in this paper, we consider completely different perturbation models and come up with a novel taxonomy. They evaluate their approach on a numerical program. Our goal is different: we want to understand the phenomenon of correctness attraction, as opposed to analyze portability.

In his PhD thesis, Khoo [5] proposes a novel kind of perturbation analysis to detect structurally equivalent program. His core hypothesis is that two programs are structurally similar if they end in the same state after the same perturbation. His perturbation technique is similar to ours but the goal is also completely different.

Tallam *et al.* [13] use three execution perturbations for run-time repair: changing thread scheduling, increasing allocated memory and denying client requests. Based on logging and checkpoint techniques, they showed that their scheme is able to avoid environment to occurs again. Compared to our work, the goal and perturbation models are different: 1) they want to perform automatic repair while we aim at understanding the perturbability of software 2) they consider coarse-grain environment perturbations while we consider fine-grain, intra-method perturbations.

In [14], they devise a framework for studying stability and instability in floating-point programs. They consider two perturbations for floating-point numbers: changing the least-significant bit, changing floating-point expressions by an "equivalent" form but syntactically different. We consider completely different perturbation models, hence our results shed a novel light on the correctness attraction.

Our work is related to approximate computing. This research area is very large and we refer to the good recent survey by Mittal *et al.* [7]. Approximate computing also modifies the execution in applications where users can accept accuracy losses in the results, e.g., in areas such as image processing or machine learning. Yet, approximate computing and correctness attraction are built on fundamentally different assumptions: the core insight of approximate computing is that a small change in the execution yields an acceptable yet degraded result. Meanwhile, correctness attraction refers to a different phenomenon: a small change in the execution yields an unchanged, fully correct result. This major distinction between both approaches is revealed in the experimental protocols: we focus on perfect oracles and we do not accept approximated or less accurate results. We note that the oracles for program laguerre and linreg are alike approximate computing, yet the error margin we accept is very low ($10^{-6}$). The work of Roy *et al.* [11] to identify approximable portions of code is related to our protocol ATTRACT. By using binary instrumentation, the value of some variables is perturbed and the new output is measured. The main differences are that: 1) they consider the classical perturbation model consisting of bitflips, second they perform a small random exploration of the perturbation space while we perform an exhaustive exploration.

To sum up, our work technically differs from previous works by the use of systematic exploration of the perturbation space and the use of perfect oracles. Our taxonomy of the causes of correctness attraction is unique and is not present in previous work.

## 8 Conclusion

We have devised a protocol called ATTRACT to study the stability of programs under perturbation. ATTRACT exhaustively explores the perturbation space of a given program for a set of inputs according to a perturbation model. We have explored the perturbability of 10 subjects for two perturbation models, PONE for integers and PBOOL for booleans. In total, 2917701 perturbed executions have been done and studied, which makes it one of the largest perturbation experiment ever made.

We have observed the presence of "correctness attraction" in all of them: 67.76% of perturbations do not break the correctness of the output. Our manual analysis of perturbed yet correct executions yields an original taxonomy of the causes behind correctness attraction. This taxonomy provides a foundation to future work that will explicitly engineer, if not maximize, correctness attraction.

## References

1. E. Barr, M. Harman, P. McMinn, M. Shahbaz, and S. Yoo. The oracle problem in software testing: A survey. *IEEE Transactions on Software Engineering*, 41(5):507–525, May 2015.
2. B. Baudry and M. Monperrus. The Multiple Facets of Software Diversity: Recent Developments in Year 2000 and Beyond. *ACM Computing Surveys*, pages 1–26, 2015.
3. E. W. Dijkstra. On the cruelty of really teaching computing science. Dec. 1988.
4. P. R. Eggert and D. S. Parker. Perturbing and evaluating numerical programs without recompilation—the wonglediff way. *Software: Practice and Experience*, 35(4):313–322, 2005.
5. W. M. Khoo. *Decompilation as Search.* PhD thesis, University of Cambridge, 2013.
6. X. Li and D. Yeung. Application-level correctness and its impact on fault tolerance. In *2007 IEEE 13th International Symposium on High Performance Computer Architecture*, pages 181–192, Feb 2007.
7. S. Mittal. A survey of techniques for approximate computing. *ACM Comput. Surv.*, 48(4):62:1–62:33, Mar. 2016.
8. L. Morell, B. Murrill, and R. Rand. Perturbation analysis of computer programs. In *Computer Assurance, 1997. COMPASS '97. Are We Making Progress Towards Computer Assurance? Proceedings of the 12th Annual Conference on*, pages 77–87, Jun 1997.

9. R. Pawlak, M. Monperrus, N. Petitprez, C. Noguera, and L. Seinturier. Spoon: A Library for Implementing Analyses and Transformations of Java Source Code. *Software: Practice and Experience*, 46:1155–1179, 2015.
10. M. Rinard, C. Cadar, and H. H. Nguyen. Exploring the acceptability envelope. In *Companion to the 20th Annual ACM SIGPLAN Conference on Object-oriented Programming, Systems, Languages, and Applications*, OOPSLA '05, pages 21–30, New York, NY, USA, 2005. ACM.
11. P. Roy, R. Ray, C. Wang, and W. F. Wong. Asac: Automatic sensitivity analysis for approximate computing. *SIGPLAN Not.*, 49(5):95–104, June 2014.
12. R. Sedgewick. Implementing quicksort programs. *Communications of the ACM*, 21(10):847–857, 1978.
13. S. Tallam, C. Tian, R. Gupta, and X. Zhang. Avoiding program failures through safe execution perturbations. In *Proceedings of the 2008 32Nd Annual IEEE International Computer Software and Applications Conference*, COMPSAC '08, pages 152–159, Washington, DC, USA, 2008. IEEE Computer Society.
14. E. Tang, E. Barr, X. Li, and Z. Su. Perturbing numerical calculations for statistical analysis of floating-point program (in)stability. In *Proceedings of the 19th International Symposium on Software Testing and Analysis*, ISSTA '10, pages 131–142, New York, NY, USA, 2010. ACM.
15. N. Wang, M. Fertig, and S. Patel. Y-branches: when you come to a fork in the road, take it. In *12th International Conference on Parallel Architectures and Compilation Techniques*, pages 56–66, Sept 2003.
16. T. A. Welch. A technique for high-performance data compression. *Computer*, 17(6):8–19, June 1984.

## A Experiment Subject

### A.1 Overview

| Subject | LOC | $N_{pp}^{int}$ | $N_{pp}^{bool}$ | $N_{pp}$ | description |
|---|---|---|---|---|---|
| quicksort | 42 | 41 | 6 | 47 | sort an array of integer |
| zip | 56 | 19 | 6 | 25 | compress a string without loss |
| sudoku | 87 | 89 | 26 | 115 | solve 9x9 sudoku grid |
| md5 | 91 | 164 | 10 | 174 | hash a message 5 |
| rsa | 281 | 117 | 20 | 137 | crypt with public and private keys |
| rc4 | 146 | 115 | 7 | 122 | crypt with symmetric key |
| canny | 568 | 450 | 79 | 529 | edge detector |
| lcs | 43 | 79 | 9 | 88 | compute the longuest common sequence |
| laguerre | 440 | 72 | 25 | 97 | find roots for polynomial functions |
| linreg | 188 | 75 | 15 | 90 | compute the linear regression from set of points |

Table 9: Dataset of 10 subjects programs used in our experiments

Table 9 gives an overview of the considered benchmark. The 1st column is the name used to refer to the subject and the second column gives the number of Line of Code (LOC) of the program. Then in the 3rd, 4th and 5th respectively the number of integer perturbation point, the number of Boolean perturbation point and the total number of perturbation point for each subject. In the last column, it a brief description the computation considered.

### A.2 Quicksort

Quicksort is a sorting algorithm. We consider an implementation of Quicksort algorithm in Java. The original code is available at `https://frama.link/XGMAr134`. A live demo is available at `https://danglotb.github.io/resources/correctness-attraction/live-demo.html`

| Subject | Oracle | Input Generator |
|---|---|---|
| quicksort | https://frama.link/oracle-qs | https://frama.link/input-qs |
| zip | https://frama.link/oracle-zip | https://frama.link/input-zip |
| sudoku | https://frama.link/oracle-sudoku | https://frama.link/input-sudoku |
| md5 | https://frama.link/oracle-md5 | https://frama.link/input-md5 |
| rsa | https://frama.link/oracle-rsa | https://frama.link/input-rsa |
| rc4 | https://frama.link/oracle-rc4 | https://frama.link/input-rc4 |
| canny | https://frama.link/oracle-canny | https://frama.link/input-canny |
| lcs | https://frama.link/oracle-lcs | https://frama.link/input-lcs |
| laguerre | https://frama.link/oracle-laguerre | https://frama.link/input-laguerre |
| linreg | https://frama.link/oracle-linreg | https://frama.link/input-linreg |

Table 10: URL to oracle and input generator for each of the 10 subject

*Correctness Oracle:* The oracle checks that the array is correctly sorted, checks that each element of the input is also in the output, and checks that no element that is not present in the input is in the output.

### A.3 Zip

The Lempel-Ziv-Welch (LZW) [16] is a loss-less data compression algorithm. We use it to compress/uncompress strings. The implementation comes from Rosetta Code[3], with 1 class and 2 methods: one class to compress, and the other class to uncompress. The implementation has 6 Boolean perturbation points and 19 numerical perturbation points spread over 56 lines of code.
*Correctness Oracle:* The scenario is to uncompress the compressed input string. The perfect oracle asserts that the output string is the same as the input string.

### A.4 Sudoku

We consider a Sudoku solver taken from Rosetta Code. We input a randomly generated grid. Some cells are already filled in with values. There is 1 class of 87 lines of codes, containing 89 numerical perturbation points and 26 Boolean perturbation points.
*Correctness Oracle:* The oracle asserts that all Sudoku constraints are satisfied: all cells are filled and valid, and all cells already in the input problem remain unchanged.

### A.5 MD5

The Message Digest 5 (MD5) algorithm is used to hash a string of a given size. We take the implementation from Rosetta Code. There is 1 class with 1 method, and 91 lines of codes. We find 164 numerical perturbation points, and 11 Boolean perturbation points.
*Correctness Oracle:* The oracle is that the hash is the same as the one from the reference implementation.

### A.6 RSA

An RSA cryptosystem was designed by Ron Rivest, Adi Shamir, and Leonard Adleman. This implementation is a real, production-ready one taken from bouncy-castle[4][5]. The project is composed of 1494 classes with a total of 241483 lines of code. We studied the RSACoreEngine class, which has 6 methods with 203 lines of codes, 73 numerical perturbation points and 19 Boolean perturbation points. Many integer points are BigInteger Java objects, that we perturb appropriately. The considered inputs are random strings of 64 bytes. *Correctness Oracle:* The considered scenario is decrypt(crypt(x)): The oracle asserts that the decrypted string is the same as the input string.

[3] http://rosettacode.org/
[4] https://www.bouncycastle.org/
[5] https://github.com/bcgit/bc-java

### A.7 RC4

RC4 is an encryption cipher designed by Ron Rivest. This algorithm is fast and simple yet not secure according to today's standards. We use BouncyCastle's class RC4CoreEngine which has 150 lines with 7 Boolean perturbation points and 112 integer points.
*Correctness Oracle:* The considered scenario is decrypt(crypt(x)). The oracle asserts that the decrypted string is the same as the input string.

### A.8 Canny

A canny filter is an edge detector in an image. We use the implementation of Tom Gibara[6]. There is one 1 class with 568 lines of code, with 450 integer perturbation points and 79 Boolean perturbation points.
*Correctness Oracle:* The oracle asserts that the detected edges are accurate of to the pixel with regards to the result of an unperturbed reference run.

### A.9 LCS

We consider the Longest Common Sequence problem, implemented using dynamic programming [7]. As input, we use real RNA sequences of two plants: *sativa* and *thaliana*, extracted from the mature dataset of miRBase[8]. This implementation has 43 Lines with 9 Boolean perturbations point and 79 integer perturbation points.
*Correctness Oracle:* The oracle is that the output is the same as the one of the reference unperturbed implementation.

### A.10 Laguerre

Laguerre is an numerical analysis program which computes the the roots of a polynomial equation. The implementation comes from The Apache Commons Mathematics Library [9]. The class under study is "LaguerreSolver" which is 440 lines long and has 176 interger perturbation points and 25 Boolean perturbation points.
*Correctness Oracle:* The oracle checks if the computed solution actually nullifies the equation. Because the computation acts on floating-point numbers, we accept the solution if its evaluation is within $+/-10^{-6}$.

[6] http://www.tomgibara.com/computer-vision/canny-edge-detector
[7] https://frama.link/3ZxP5eBj
[8] http://www.mirbase.org/ftp.shtml
[9] version 3.6.1: https://frama.link/tQCYrZ2W

## A.11 Linreg

Linreg computes a linear regression using the Tikhonov regularization. We take the implementation from the Weka Library [10]. The class under study is "LinearRegression": it has 188 lines of codes, with 75 integer perturbation points and 15 Boolean perturbation points. We generate inputs by randomly sampling the coefficients of the equation.
*Correctness Oracle:* It checks if the computed coefficients are equal to those obtained from a reference run, up to a $10^{-6}$ precision.

[10] version 3.8.0: https://frama.link/fCjiqzk2